\documentclass[10pt,letterpaper]{article}

\begin{document}

\title{No minimally coupled scalar black hole hair in Lanczos-Lovelock gravity}

\author{Jozef Sk\'akala\footnote{jozef@iisertvm.ac.in}~ and
S. Shankaranarayanan \footnote{shanki@iisertvm.ac.in} \\ Indian Institute of Science, Education and Research (IISER-TVM),\\ Trivandrum 695016, India}

\maketitle

\begin{abstract}
We extend here the result of Bekenstein \cite{Bekenstein1, Bekenstein2} proving the non-existence of minimally coupled scalar black hole hair in general relativity to the Lanczos-Lovelock gravity in arbitrary dimension with non-negative coupling constants. The only physical requirement on the multiplet of minimally coupled scalar fields is that it fulfills the weak energy condition. We also assume, similarly to Bekenstein, spherical symmetry and asymptotic flatness.
\end{abstract}

\maketitle

\section{Introduction}

The no hair theorems in the black hole physics have a long and interesting history \cite{Robinson}. The most important ``no hair'' results focused on proving the uniqueness of the Kerr-Newman metric in the electro-vacuum. However another important line of research focused on proving the non-existence of other independent black hole's quantum numbers beyond mass, angular momentum and charge, quantum numbers associated with the fields in the black hole exterior. In this context it is certainly interesting to explore if a black hole can be endowed with a minimally coupled scalar hair. This question was investigated some time ago in the context of general relativity \cite{Bekenstein1, Bekenstein2} and also long time ago in a more general context \cite{Bekenstein2, Bekenstein3}. (For non-minimally coupled scalar hair see e.g. \cite{Bekenstein4, Saa1, Saa2}.)

Let us now briefly summarize some of the basic results regarding minimally coupled scalar hair: Take a minimally coupled self-interacting scalar field with the action 
\[S=-\frac{1}{2}\int d^{D}x \sqrt{-g}(\psi_{,i}\psi^{,i}+V(\psi^{2})).\]
Bekenstein had shown \cite{Bekenstein2, Bekenstein3}, that if the scalar field potential everywhere fulfills the condition $V_{,\psi^{2}}\geq 0$, where the lower case index stands for the derivative with respect to the field variable squared, then by using only the wave equation together with some mild assumptions (like asymptotic flatness and finiteness of the stress energy tensor invariants at the horizon) one can prove that a stationary non-trivial scalar field cannot form around a black hole. Since this proof uses only the scalar field wave equation it counts as a proof for arbitrary theory of gravity. 

However the condition on the field's potential ($V_{,\psi^{2}}\geq 0$) is quite restrictive. Even the Higgs field potential does not fulfill such a condition. Therefore Bekenstein partially extended this result \cite{Bekenstein1, Bekenstein2} to a \emph{general} minimally coupled scalar field fulfilling the weak energy condition. Moreover this result can be easily generalized to a multiplet of scalar fields \cite{Bekenstein1}. However, the extension was achieved ``only'' partially since the extended result uses the spherical symmetry condition and, more importantly, it uses also the equations of the general theory of relativity. This means the second result, valid for a more general scalar field depends, unlike the previous one, on the theory of gravity. 

In this work we want to extend the second result of Bekenstein \cite{Bekenstein1, Bekenstein2} to Lanczos-Lovelock gravity theories. (Our paper thus falls in the line of research that occurred in the last two decades and is looking at generalizations of the older no hair results to the higher-dimensional black holes, e.g. \cite{Robinson, Gibbons1, Gibbons2, Rogatko}. Another interesting line of research explores what happens with black hole uniqueness if one looks to some other, not necessarily higher-dimensional, generalizations of Einstein gravity. One very powerful result proves that in scalar-tensor gravity theories black hole uniqueness in vacuum is the same as in General Relativity \cite{Sotiriou}.) We prove here that the no general-minimally-coupled scalar hair result in spherically symmetric spacetime obtained for general relativity, (scalar field fulfilling the weak energy condition), extends (at least) to a significant subclass of Lanczos-Lovelock theories. The proof is contained in the next section of the paper and at the end of the paper we discuss our results. We use the signature convention $(-,+,...,+)$.

\section{No minimally coupled scalar hair for Lanczos-Lovelock spherically symmetric black holes}

The Lanczos-Lovelock gravity \cite{Lanczos, Lovelock} is defined by an action of the form:
\[S_{L}=\int d^{D}x\sqrt{-g}\sum_{p=0}^{[D/2]}\alpha_{p}L_{p},\]
where the Lagrangians $L_{p}$ are defined by:
\[L_{p}=\frac{1}{2^{p}}\delta^{a_{1}...a_{p}b_{1}...b_{p}}_{c_{1}...c_{p}d_{1}...d_{p}}
R^{c_{1}d_{1}}_{~~~~a_{1}b_{1}}...R^{c_{p}d_{p}}_{~~~~a_{p}b_{p}}.\]
It represents a most natural higher-dimensional extension of the general relativity theory. (For a recent review see \cite{Paddy}.) The Lanczos-Lovelock theory of gravity is for the case of our theorem constrained by the following conditions:
\begin{itemize}
\item All the coupling constants $\alpha_{p}$ in Lanczos-Lovelock theory are non-negative, $\alpha_{p}\geq 0$. Let us mention that in the case of Gauss-Bonnet gravity (second order Lanczos-Lovelock theory) it is known that $\alpha_{2}<0$ leads to ghosts \cite{Boulware}. Hence only positive coupling constant is allowed.
\item We assume the theory includes general relativity, therefore $\alpha_{1}\neq 0$.
\item Spacetime is asymptotically flat, which is equivalent to the condition $\alpha_{0}=0$, as shown in Ref.\cite{Maeda}. (This means we exclude the cosmological constant.)
\end{itemize} 
We also need to mention that, as Bekenstein in his original proof, we are constraining the spacetime by the condition of spherical symmetry.

This means, take the $D$ dimensional spherically symmetric static spacetime with the metric:
\[-f(r)dt^{2}+g^{-1}(r)dr^{2}+r^{2}\gamma_{ij}dx^{i}dx^{j},~~~i,j=2,3...D-1,\]
where $\gamma$ is the metric of $D-2$ dimensional sphere. The line element is supposed to represent a black hole in the asymptotically flat spacetime. Both $f$ and $g$ are everywhere positive in the black hole exterior and the function $f$ must vanish at the black hole horizon $r_{H}$. Moreover, by calculating the scalar curvature of such a geometry it can be easily observed that $f(r_{H})=0$ and $g(r_{H})\neq 0$ implies curvature singularity at the horizon. Therefore $g$ is also required to always vanish at the horizon.

\subsection{Part of the proof that uses the scalar field equation}

Similarly to Bekenstein \cite{Bekenstein1, Bekenstein3} consider the action of a multiplet of $n=1,...,N$ scalar fields with generalized dynamics
\begin{eqnarray}\label{scaction}
S=S_{L}-\int L(\psi_{1}...\psi_{N},I_{1}...I_{N})\cdot\sqrt{-g}\cdot d^{D}x,~~~~~~~~~~~\\ 
I_{n}=\psi_{n}^{~,j}\psi_{n,j},~~~~~~~\nonumber
\end{eqnarray}
where $S_{L}$ is (this time) the action of the Lanczos-Lovelock theory. 
The stress energy tensor of the scalar field multiplet reads:
\begin{equation}\label{stress}
T^{\mu}_{\nu}=2\sum_{n=1}^{N}\frac{\partial L}{\partial I_{n}}\psi_{n,\nu}\psi_{n}^{~,\mu}-L\delta^{\mu}_{\nu}.
\end{equation}
Now one can easily observe from Eq.(\ref{stress}) that
\begin{equation}\label{t}
T^{t}_{t}=T^{i}_{i}=-L,
\end{equation}
where $T^{i}_{i}$ is arbitrary angular diagonal element of the stress energy tensor, and
\begin{equation}\label{x}
T^{r}_{r}=2g\sum_{n=1}^{N}\frac{\partial L}{\partial I_{n}}(\psi_{n,r})^{2}-L.
\end{equation}
The weak energy condition implies $T^{t}_{t}\leq 0$. Furthermore \cite{Bekenstein3}, the causality condition together with the weak energy condition imply the following:
\begin{equation}\label{asympt}
-T^{t}_{t}+T^{r}_{r}\geq 0.
\end{equation}
 We therefore require these conditions to hold in our proof.

We can further recollect part of Bekenstein's argumentation \cite{Bekenstein1, Bekenstein2}, such that uses purely the wave equation and therefore holds for any theory of gravity. The $r$-component of the wave equation $T^{\mu}_{\nu ;\mu}=0$ can be written as: 
\begin{equation}\label{WaveEq}
\left(T^{r}_{r}\right)_{,r}=-\left(\frac{D-2}{r}+\frac{f_{,r}}{2f}\right)(-T^{t}_{t}+T^{r}_{r}).
\end{equation}

The argument goes as follows:
Equation
(\ref{WaveEq}) together with Eq.(\ref{asympt}) means that $T^{r}_{r,r}\leq 0$ near the horizon. (The surface gravity of the black hole horizon is for a non-extremal black hole always positive, therefore $f_{,r}> 0$ near the horizon. For the extremal black hole $f_{,r}=0$, but as $f$ is in the black hole exterior positive and at the horizon zero, there must still exist a neighborhood of the horizon where $f_{,r}\geq 0$.)  
Furthermore Eq.(\ref{WaveEq}) can be expressed as:
\begin{equation}\label{WaveEq2}
(r^{D-2}\sqrt{f}\cdot T^{r}_{r})_{,r}=(r^{D-2}\sqrt{f})_{,r}T^{t}_{t}
\end{equation}
and integrating leads to
\begin{equation}\label{int}
T^{r}_{r}=\frac{1}{r^{D-2}\sqrt{f}}\int_{r_{H}}^{r}(\tilde r^{D-2}\sqrt{f})_{,\tilde r}\cdot T^{t}_{t}d\tilde r.
\end{equation}
(The integration constant in Eq.(\ref{int}) is chosen to avoid the divergence of radial pressure at the horizon.) Eq.(\ref{int}) tells us that, since $f$ is an increasing function near the horizon, $T^{r}_{r}\leq 0$ near the horizon.

Furthermore Eq.(\ref{asympt}) and Eq.(\ref{WaveEq}) mean that at the asymptotic infinity $T^{r}_{r,r}\leq 0$. (At the asymptotic infinity the fact that gravity attracts matter to the gravitating body means $f_{,r}>0$. This statement, which in GR corresponds to the positivity of ADM mass will be assumed in the proof, but it will be \emph{independently} shown to hold at the end of the section.) At the same time $T^{r}_{r}$ has to vanish at infinity. Therefore the radial pressure near the horizon is non-positive and non-increasing and near the asymptotic infinity non-negative and non-increasing. 

This means, considering a non-trivial field configuration, there must be an interval where the radial pressure is increasing, which means at the interval it holds that $T^{r}_{r,r}>0$. From Eq.(\ref{WaveEq}) we see that this is possible only if 
\begin{equation}\label{point}
f_{,r}<-2f\cdot \frac{D-2}{r}<0.
\end{equation} 
Furthermore, since at the asymptotic infinity $T^{r}_{r}\geq 0$ holds, there must exist a point $r_{b}$ at which $T^{r}_{r}(r_{b})=0$ and in the same time on one of the intervals $(r_{b},r_{c})$, or $(r_{a}, r_{b})$~ $T^{r}_{r}$ is increasing. However this fact and Eq.(\ref{point}) imply that there must exist an interval on which simultaneously $T^{r}_{r}\geq 0$ and $f_{,r}<0$.

This is all one can obtain purely by using the wave equation, in order to push the argument further one has to look at a particular theory of gravity. 
Let us therefore demonstrate by using the Lanczos-Lovelock equations (and therefore generalize Bekenstein's argument to the Lanczos-Lovelock theory), that the conditions $f_{,r}<0$ and $T^{r}_{r}\geq 0$ cannot be simultaneously fulfilled for non-negative coupling constants. 

\subsection{Part of the proof that uses the gravity equations}

Let us write the radial Lanczos-Lovelock equation $\mathcal{G}^{r}_{r}=8\pi G T^{r}_{r}$ as:

\begin{eqnarray}\label{Lovelock1}
\beta_{1}\cdot \frac{g}{f}\cdot f_{,r}-\beta_{2}=8\pi G T^{r}_{r},~~~~~~
\end{eqnarray}
where 
\begin{eqnarray}\label{beta1}
\beta_{1}=\sum_{p=1}^{[D/2]}\alpha_{p}\left\{p(D-2)\cdot\frac{1}{2r}\left(\frac{1-g}{r^{2}}\right)^{p-1}\right\},
\end{eqnarray}
and
\begin{eqnarray}\label{beta2}
\beta_{2}=\sum_{p=1}^{[D/2]}\alpha_{p}\left\{\frac{(D-2)(D-2p-1)}{2}\left(\frac{1-g}{r^{2}}\right)^{p}\right\}.~~~~~~
\end{eqnarray}
~\\
Now from Eq.(\ref{Lovelock1}) one can easily see that, for non-negative coupling constants, if 
\[(1-g)> 0\]
holds everywhere, then $T^{r}_{r}\geq 0$ implies $f_{,r}> 0$.
This is because $f_{,r}$ can be expressed using Eq.(\ref{Lovelock1}) as
\[f_{,r}=\frac{f}{g}\cdot\frac{8\pi G T^{r}_{r}+\beta_{2}}{\beta_{1}},\]
where $\beta_{1}, \beta_{2}$ are positive if $(1-g)$ is positive.

This means all that remains is to show that $(1-g)$ is necessarily positive everywhere in the black hole exterior. To show this we will use the definition of Misner-Sharp mass together with the equation for the Misner-Sharp mass radial derivative. (Both of the equations can be found in \cite{Maeda}.)

The Misner-Sharp mass reads \cite{Maeda} as:
  
\begin{eqnarray}\label{MS}
M(r)=
\frac{(D-2)V_{D-2}}{16\pi G}\sum_{p=1}^{[D/2]}\alpha_{p}\frac{(D-3)!(D-2p)}{(D-2p)!}r^{D-1-2p}(1-g)^{p},
\end{eqnarray}
and the radial derivative of the Misner-Sharp mass follows the equation \cite{Maeda}:
\begin{equation}\label{MSr}
M_{,r}=-V_{D-2}T^{t}_{t}\cdot r^{D-2}.
\end{equation}

($V_{D-2}$ is the volume of the $(D-2)$-dimensional sphere.) Eq.(\ref{MSr}) means that Misner-Sharp mass is everywhere a non-decreasing function of $r$. Therefore if $M$ was negative at infinity (which violates the mass positivity theorem), it also has to be negative at the horizon. In case $M$ is non-negative at infinity it would have to become non-positive at the horizon, unless $(1-g)> 0$ everywhere in the black hole exterior.
The reason for the last statement is the following: Near the asymptotic infinity $(1-g)$ is infinitesimaly small, hence the dominant term in Eq.(\ref{MS}) is $p=1$ (the general relativistic term). This is because the term contains the lowest power in $(1-g)$ and simultaneously the highest power in $r$ (which diverges at infinity). 

Since the power of $(1-g)$ is one, the positivity of mass means that $(1-g)>0$ near infinity. However if $(1-g)$ turned somewhere in the black hole exterior negative, as can be seen from Eq.(\ref{MS}), $M$ would become zero at the same point at which $(1-g)$ becomes zero. However this implies that $M$ would become non-positive at the horizon.

In any case, unless $(1-g)> 0$ everywhere in the black hole exterior, the Misner-Sharp mass would become non-positive at the horizon. Non-positivity of the Misner-Sharp mass however contradicts the existence of the horizon, as the horizon's radius $r_{H}$ is given by Eq.(\ref{MS}) with $g=0$. This gives:
 
\begin{eqnarray}\label{horizon}
M(r_{H})=
\frac{(D-2)V_{D-2}}{16\pi G}\sum_{p=1}^{[D/2]}\alpha_{p}\frac{(D-3)!(D-2p)}{(D-2p)!}r_{H}^{D-1-2p}.
\end{eqnarray}

However since the coefficients on the right side of equation (\ref{horizon}) are all positive, equation (\ref{horizon}) will have no positive real solution $r_{H}$ for non-positive $M$. This proves that $(1-g)>0$ everywhere in the black hole exterior. As we have shown this implies that $f_{,r}<0$ and $T^{r}_{r}>0$ cannot hold in the same time. As shown before, this was the statement we needed in order to complete the proof that there is no minimally coupled spherically symmetric black hole scalar hair in case of Lanczos-Lovelock gravity with non-negative coupling constants.  

There is one remaining detail that needs to be shown: Throughout the proof we assumed as obvious the fact that $f_{,r}>0$ at the asymptotic infinity. However this statement follows from the fact that $(1-g)> 0$ everywhere in the black hole exterior and $g\to 1$ at the asymptotic infinity. (These two statements are independent on the assumption of $f_{,r}>0$ at the infinity, hence the proof is not circular.) This behaviour of $g$ implies $g_{,r}>0$ at the asymptotic infinity. Let us write now the following combination of Lanczos-Lovelock equations: $-\mathcal{G}^{t}_{t}+\mathcal{G}^{r}_{r}=8\pi G(-T^{t}_{t}+T^{r}_{r})$. This reads as:

\begin{eqnarray}\label{Lovelock1b}
\left(\frac{g f_{,r}}{f}-g_{,r}\right)\beta_{1}=8\pi G (-T^{t}_{t}+T^{r}_{r})\geq 0.~~~~
\end{eqnarray}
In the inequality in Eq.(\ref{Lovelock1b}) we used Eq.(\ref{asympt}). But this means, considering the fact that $(1-g)>0$, that
\[f_{,r}\geq\frac{f}{g}g_{,r}>0,\]
at asymptotic infinity.

\section{Discussion}

In this work we have shown that Bekenstein's no hair result \cite{Bekenstein1, Bekenstein2} for a general multiplet of minimally coupled scalar fields fulfilling the weak energy condition generalizes to Lanczos-Lovelock theory of gravity with non-negative coupling constants.

Let us further make two remarks:
Firstly the fact that the stress energy tensor describes a scalar field given by the action (\ref{scaction}) was used only in one single step. It was used in the fact that $T^{t}_{t}=T^{i}_{i}$, where by $T^{i}_{i}$ we mean the diagonal angular component of the stress-energy tensor. (All the diagonal angular components of the stress-energy tensor are in case of spherical symmetry necessarily equal.) This means the proof given here for scalar fields holds automatically for any other matter field described by a stress energy tensor that fulfills the condition $T^{t}_{t}=T^{i}_{i}$.  

Secondly, equation (\ref{Lovelock1}) can be expressed more generally for a spacetime containing a maximally symmetric $D-2$ dimensional subspace as

\begin{eqnarray}\label{Lovelock2}
\sum_{p=1}^{[D/2]}\alpha_{p}\left\{p(D-2)\cdot\frac{g f_{,r}}{2rf}\left(\frac{k-g}{r^{2}}\right)^{p-1}-\right.~~~~~~~~~~~~~~~~~~~~~~~~~~~~~~~~~~\nonumber\\
\left.\frac{(D-2)(D-2p-1)}{2}\left(\frac{k-g}{r^{2}}\right)^{p}\right\}=8\pi G T^{r}_{r}.~~~~~~
\end{eqnarray}

Here $k=\{\pm 1,0\}$ is a sectional curvature of the maximally symmetric subspace (sphere, hyperboloid, plane). One can easily see that Eq.(\ref{Lovelock2}) and $T^{r}_{r}>0$ directly imply that $f_{,r}>0$ for two subcases, which are however not of any substantial interest. For a general spacetime with maximally symmetric subspace in a theory with $\alpha_{p}\geq 0$ where $\alpha_{p}=0$ for $p$ odd. Also the same implication one obtains for the case of $k=-1,0$ in the theory with $\alpha_{p}\leq 0$ where $\alpha_{p}=0$ for $p$ even. Therefore the proof of no scalar hair simply holds also for these cases. (Note that the wave equation part of the argument trivially generalizes from spherically symmetric spacetime to the more general spacetime with maximally symmetric $D-2$ dimensional subspace.) However, it is important to note that, in general, relaxing the condition on $\alpha_{p}\geq 0$, or departing from spherical symmetry (in any direction), has significant consequences for the logic used in our proof and new ideas have to be employed. (Note also that with spinning black holes the situation is far more complicated: It is well known that vacuum rotating black holes in higher dimensional gravity have more rich structure than in 4D \cite{Emparan}. Furthermore, it has been recently shown that even a non-stationary \emph{complex massive} scalar field in ordinary 4D General Relativity can lead to a stationary, rotating, non-trivial black hole exterior \cite{Herdeiro}.)

In future work it would be interesting to look at what happens with the non-minimally coupled scalar fields. Also in case of curvature coupled field and general relativity Bekenstein had shown that the no hair result holds for certain intervals of values of the scalar field coupling parameter \cite{Bekenstein2, Bekenstein3}. However outside these intervals there is a known non-trivial (albeit unstable) solution describing a configuration of scalar field around the static black hole \cite{Bekenstein2}. It would be interesting to see to what extend these results generalize in the Lanczos-Lovelock theory. It would be also interesting to explore what happens in theories with non-metric connection (metric-affine or Palatini gravity) \cite{Cappozzielo}. All this is left for future research.

\bigskip

{\bf Acknowledgments:}~~ We would like to thank D. Kothawala for discussions. The work is supported by Max Planck partner
group in India. SS is partly supported by Ramanujan Fellowship of DST,
India.

\end{document}